\documentclass[prl,twocolumn]{revtex4-2}

\usepackage{booktabs} 
\usepackage[top=1in, bottom=1in, left=1in, right=1in]{geometry}
\usepackage{color} 
\usepackage{amsfonts,amssymb,amsmath} 
\usepackage{bm} 
\usepackage{bbm}
\usepackage{hyperref}
\usepackage{fancyvrb}
\usepackage{xr}
\usepackage{tikz}
\usepackage{tkz-euclide}
\usetikzlibrary{decorations.pathmorphing}	
\tikzset{
    v/.style={decorate, decoration={snake, segment length=3mm, amplitude=0.75mm}, draw},
    f/.style={draw,decoration={markings,mark=at position #1 with {\arrow[very thick]{latex}}},postaction={decorate},node contents=#1},
    f/.default=.6,
    fb/.style={draw,decoration={markings,mark=at position #1 with {\arrowreversed[very thick]{latex}}},postaction={decorate},node contents=#1},
    fb/.default=.4,
    fnar/.style={draw},
    g/.style={decorate, draw,  decoration={coil,amplitude=3pt, segment length=3.5pt}},
    s/.style={dashed,draw, postaction={decorate},
        decoration={markings,mark=at position .55 with {\arrow[very thick]{latex}}}},
    sb/.style={dashed,draw, postaction={decorate},
        decoration={markings,mark=at position .55 with {\arrowreversed[draw=black,very thick]{latex}}}},
    snar/.style={dashed,draw,line width =1.25pt},
}
\tikzset{every picture/.style={line width=1}}

\definecolor{c1}{rgb}{0., 0.26, 0.5}
\definecolor{c2}{rgb}{0.9,0.648, 0.}
\definecolor{c3}{rgb}{0.75, 0., 0.}
\definecolor{c4}{rgb}{0., 0.6, 0.6}
\definecolor{c5}{rgb}{0.357, 0.35, 0.7}
\definecolor{c6}{rgb}{0.588, 0.6, 0.}

\begin{document}

\title{New Opportunities for Detecting Axion-Lepton Interactions}

\author{Wolfgang Altmannshofer}
\email{waltmann@ucsc.edu} %

\author{Jeff A. Dror}
\email{jdror1@ucsc.edu} %

\author{Stefania Gori}
\email{sgori@ucsc.edu} %

\affiliation{Department of Physics, University of California Santa Cruz, 1156 High St., Santa Cruz, CA 95064, USA\\
and Santa Cruz Institute for Particle Physics, 1156 High St., Santa Cruz, CA 95064, USA
}

\date{\today}


\begin{abstract}
We revisit the theory and constraints on axion-like particles (ALPs) interacting with leptons. We clarify some subtleties in the constraints on ALP parameter space and find several new opportunities for ALP detection. We identify a qualitative difference between weak-violating and weak-preserving ALPs, which dramatically change the current constraints due to possible ``energy enhancements'' in various processes. This new understanding leads to additional opportunities for ALP detection through charged meson decays (e.g., $\pi^+\to e^+ \nu a$, $K^+\to e^+ \nu a$) and $ W $ boson decays. The new bounds impact both weak-preserving and weak-violating ALPs and have implications for the QCD axion and addressing experimental anomalies using ALPs.
 \end{abstract}

\maketitle
\section{Introduction}
Axion-like particles (ALPs) below the electroweak scale are a simple, natural extension of the Standard Model (SM). Beyond the original motivation to address the strong CP problem, they are one of the primary targets for the large-scale experimental effort to search for {\em hidden sectors}~\cite{Jaeckel:2010ni,Ringwald:2012hr,Essig:2013lka}. ALPs are characterized by their derivatively-coupled interactions, which enjoy an underlying global symmetry in the ultraviolet referred to as a Peccei-Quinn (PQ) symmetry~\cite{Peccei:1977hh}. An ALP mass is generated if the theory exhibits a small explicit breaking of this symmetry and is parametrically suppressed relative to the axion decay constant. The charges of the SM fields under the global symmetry dictate the ALP interactions. In this paper, we focus on ALPs coupled primarily to leptons. Detection of such ``leptophilic'' ALPs has been widely considered within terrestrial experiments~\cite{Konaka:1986cb,Riordan:1987aw,Bjorken:1988as,Bross:1989mp,Scherdin:1991xy,Tsai:1989vw,Bassompierre:1995kz,Izaguirre:2016dfi,Marciano:2016yhf,Berlin:2018bsc,AristizabalSierra:2020rom,Gori:2020xvq,Bauer:2020jbp,Bauer:2021mvw,Carenza:2021osu,Buen-Abad:2021fwq}, for ALPs produced in the Sun~\cite{EDELWEISS:2018tde,LUX:2017glr,PandaX:2017ock,XENON:2020rca}, and in various astrophysical environments~\cite{Isern:2008nt,MillerBertolami:2014rka,Capozzi:2020cbu,Straniero:2020iyi,VanTilburg:2020jvl,Bollig:2020xdr,Croon:2020lrf,Lucente:2021hbp}. 

In this work, we revisit the theory behind leptophilic ALPs. Our primary result is the calculation of ALP production from leptonic charged meson decays,~\footnote{Charged pion decays have previously been considered as relevant probes of hadronic and muonic ALP couplings~\cite{Krauss:1986bq,Bardeen:1986yb,Altmannshofer:2019yji,Hostert:2020xku,Bauer:2021wjo,Cheung:2022umw}.} which can be searched for at meson factories and proton beam dumps. We show that the decay rates depend critically on the symmetry structure of the coupling. If the charges of the underlying PQ symmetry commute with those of the SM weak interaction, then searches for ALPs produced in $ \pi ^\pm $ and $ K ^\pm $ decays can lead to limits on leptonic couplings a factor of a few more stringent than present bounds and even constrain open parameter space of the QCD axion. If the global symmetry charges do not commute with the SM weak interactions, meson decay rates are further enhanced by factors of $ ({\rm energy}/m _\ell) ^2 $, where $ m _\ell $ is the lepton mass.~\footnote{The energy enhancement is closely analogous to that discovered for light gauge bosons coupled to axial currents~\cite{Dror:2017ehi} and differences in lepton numbers \cite{Dror:2020fbh} (see also Refs.~\cite{Fayet:2006sp, Barger:2011mt, Karshenboim:2014tka,Dror:2017nsg,Dror:2018wfl}).} This can lead to even more stringent bounds from a number of charged meson decay channels. A similar enhancement is present in $ W $ boson decays and, in combination, our new bounds dramatically shape the leptophilic-ALP parameter space. 

\section{Theory of ALP-lepton Interactions} 
\label{sec:theory}
The structure of the PQ symmetry demands that the leading ALP interactions vanish when the lepton current, $ j ^\mu _{{\rm PQ}}    $, is conserved. This restricts the Lagrangian to the form ${\cal L} = \partial _\mu  a ~j ^\mu _{ \rm PQ}$. We consider a current that contains all three possible lepton couplings independently, 
\begin{equation} \label{eq:Jint}
j ^\mu _{ {\rm PQ}}  = \frac{\bar g_{\ell \ell }}{2m_\ell } \bar \ell  \gamma^\mu \ell  + \frac{g_{\ell \ell }}{2m_\ell } \bar \ell  \gamma^\mu \gamma_5 \ell  + \frac{g_{\nu _\ell }}{2m_\ell } \bar\nu _\ell \gamma^\mu P_L \nu _\ell \,. 
\end{equation} 
Since our aim is to set robust bounds, we focus on a single lepton flavor though it is straightforward to generalize the discussion to multiple flavors. We have chosen to normalize the dimensionful scale to the charged lepton mass, such that the couplings $g_{\ell\ell},\bar g_{\ell\ell},g_{\nu_\ell}$ are dimensionless. We do not impose electroweak invariance in Eq.~\ref{eq:Jint}, which would require $g_{\nu _\ell } = \bar g_{\ell \ell } - g_{\ell \ell }$. This condition can be violated in various ways. For example, an ALP-photon coupling generates via renormalization-group flow below the electroweak scale only $g_{\ell\ell}$ but not $\bar g_{\ell\ell}$ nor $g_\nu$. In practice, such a violation of $SU(2)$ invariance is only modest in size, unless the ALP-photon coupling is so large as to dominate the ALP phenomenology. More generally, each lepton coupling can arise independently in a weak-invariant theory by including the currents,~\footnote{Possible UV completions of the higher dimensional operators involve an axion coupling to vector-like leptons that connect to the SM leptons through Yukawa interactions. We present two examples in the supplementary material. While these models are constrained by searches for vector-like leptons~\cite{PhysRevD.97.075016,Bolton:2019pcu,Batell:2022dpx}, these limits are often weaker than the ones we derive here.}
\begin{equation} 
 \overline{ ( H L )} \gamma _\mu ( H L ) \,,  \overline{ e} _R \gamma _\mu  e _R  \,, ~{\rm and}~~  \overline{ ( H ^\dagger  L )} \gamma _\mu (  H ^\dagger   L )\,,
\end{equation} 
where $ L = ( \ell , \nu _\ell ) $ is the SU(2)-doublet and $H $ denotes the SM Higgs boson.

The impact of the leptophilic-ALP interactions is best understood by integrating the Lagrangian by parts and writing it as an axion multiplying the divergence of the PQ current. There are three types of contributions to the divergence to leading order in the axion couplings: 
\begin{align}
\partial _\mu & j _{ {\rm PQ}} ^\mu   =  g_{\ell \ell} (\bar \ell i\gamma_5 \ell)  \label{eq:int} \\ 
& +  \frac{ e ^2 }{ 16 \pi ^2 m _\ell } \bigg[  \frac{ \bar{g} _{ \ell \ell } - g _{ \ell \ell} + g _{\nu_\ell}  }{ 4 s _W ^2}  W ^+ _{\mu\nu} \tilde W ^{ - , \mu \nu } \notag \\ 
&    + \frac{\bar{g} _{ \ell \ell } - g _{ \ell \ell} ( 1 - 4  s _W ^2 ) }{ 2 c _W s _W } F _{\mu\nu} \tilde{Z} ^{\mu\nu} - g _{ \ell \ell }  F _{\mu\nu} \tilde{F} ^{\mu\nu} + \notag \\ 
& \frac{ \bar{g} _{ \ell \ell } ( 1 - 4 s _W ^2 ) - g _{ \ell \ell } ( 1 - 4 s _W ^2 + 8 s _W ^4 )  + g _{\nu _\ell } }{ 8 s _W ^2 c _W ^2 }     Z _{\mu\nu} \tilde{Z} ^{\mu\nu}     \bigg]  \notag \\ 
& + \frac{ig}{2\sqrt{2} m _\ell }(g_{\ell\ell} - \bar g_{\ell\ell} + g_{\nu _\ell })  (\bar \ell \gamma^\mu P _L \nu) W_\mu^- ~+~\text{h.c.}  \,, \notag 
\end{align}
where $s_W $ is the sine of the weak mixing angle.

The first, Yukawa-like, term drives most of ALP phenomenology and has been extensively studied in the literature. This coupling is often wrongly stated as the only physical term. This claim is only valid if one neglects the weak interactions that render each coupling in Eq.~\eqref{eq:Jint} physical. 

The second set of terms is generated through the chiral anomaly and has been studied in the context of flavor-changing meson decays into ALPs (see, e.g., Refs.~\cite{Izaguirre:2016dfi,Gori:2020xvq,Bauer:2020jbp,Bauer:2021mvw}). In UV complete ALP models, the anomaly terms containing weak gauge bosons have additional contributions from the shift of the measure under a chiral rotation and studying any process using these terms is sensitive to the specific UV completion~\cite{Quevillon:2019zrd,Bonnefoy:2020gyh,Quevillon:2021sfz}. 

The final contribution to the divergence can be used to set stringent bounds on ALPs, but has received far less attention in the literature.~\footnote{This additional term was noted in the context of stellar bounds on ALPs~\cite{Raffelt:1987yt} (see their footnote~[11]).} It is a key component of our work. The corresponding 4-point interaction term of the ALP with a lepton, neutrino, and $W$ boson is absent if the current respects electroweak symmetry. As we will show, in its presence, there is an $ ({\rm energy} / m _\ell $) enhancement in amplitudes involving $ W $ bosons. Note that the interactions of neutral gauge bosons conserve at the classical level separately left-handed charged lepton number, right-handed charged lepton number, and neutrino number and thus there are no analogous terms with photons or $Z$ bosons. Also note that, up to anomaly terms, there are only two physical couplings in the leptophilic ALP Lagrangian. This is a consequence of lepton number conservation at the classical level.

\section{Leptonic meson decays}
\label{sec:semileptonic}
\begin{figure} 
\vspace{.5cm}
\includegraphics[width=\columnwidth]{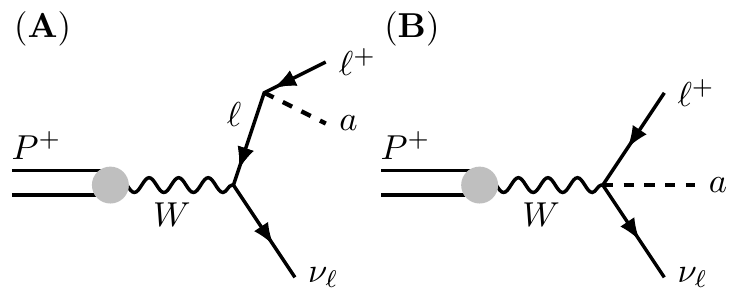}
\caption{Diagrams for the decay of a charged meson ($ P ^+ $) to an ALP ($ a $), lepton ($ \ell $), and neutrino ($ \nu _\ell  $) using the interactions in Eq.\eqref{eq:int}. ({\bf A}) relies on the standard $ \bar\ell \gamma _5 \ell $ vertex. ({\bf B}) uses the enhanced weak-violating vertex.
}
\label{fig:NPdecay}
\end{figure}
The interactions in Eq.~\eqref{eq:int} imply that there are unappreciated experimental avenues capable of probing leptophilic ALPs.
One is the charged meson ($ P ^\pm $) decay $P^\pm \rightarrow \ell^\pm \nu _\ell  a $ with the two leading order diagrams depicted in Fig.~\ref{fig:NPdecay}.
Comparing the coefficients in front of the $ \bar{\ell} \gamma _5 \ell $ and $ (\bar{\ell} \gamma ^\mu P _L \nu _\ell ) W _\mu ^- $ terms in Eq.~\eqref{eq:int}, we expect that if $ m _\ell / m _P \ll 1 $, diagram ({\bf B}) will generically be the dominant contribution to the decay rate. In a weak-preserving theory, diagram ({\bf B}) vanishes, and only diagram ({\bf A}) will contribute. In either case, the bounds from charged meson decays are a powerful probe of the leptophilic-ALP parameter space.

To present the result of our calculation, we normalize the $P ^+  \to \ell ^+  \nu _\ell a$ branching ratio to a well-measured leptonic decay of the same charged meson $P ^+  \to \ell^{\prime +} \nu _{\ell'}  $. In the ratio, all CKM matrix elements and meson decay constants cancel. In the limit $m_\ell \ll m_P$, we find
\begin{multline} \label{eq:BRpienua}
 \frac{\mathcal{B}(P ^+  \to \ell ^+ \nu _\ell  a)}{\mathcal{B}(P ^+  \to \ell^{\prime +} \nu _{ \ell ' })} \simeq \frac{1}{1536 \pi^2} \frac{m_P^4}{m_\ell^2 m_{\ell^\prime}^2} \left(1 - \frac{m_{\ell^\prime}^2}{m_P^2}\right)^{-2} \\
 \times \Bigg[ (g_{\ell\ell} - \bar g_{\ell\ell} + g_{\nu_\ell})^2 f_0(x_P)
 + \frac{16 m_\ell^2}{m_P^2} g_{\ell\ell}^2 f_1(x_P) \Bigg] .
\end{multline}
The phase space functions
\begin{eqnarray}
 f_0(x) &=& 1 - 8x + 8x^3 - x^4 - 12x^2 \log x~, \\
 f_1(x) &=& 1 + 9x - 9x^2 - x^3 + 6x(1+x) \log x ~,
\end{eqnarray}
depend on the ratio of ALP and meson masses $x_P = m_a^2/m_P^2$. In Eq.~\eqref{eq:BRpienua}, we have omitted additional terms of order $m_\ell^2/m_P^2$ proportional to the coupling combination $g_{\ell\ell} - \bar g_{\ell\ell} + g_{\nu_\ell}$ as they are negligible compared to the leading unsuppressed term. In the supplemental material, we give the full double differential branching ratio as a function of the lepton and ALP energies, which can be integrated numerically if $m_\ell$ is of the same order as $m_P$. 

As expected, the term enhanced by $ m_P^2 / m_\ell^2 $ is absent in the weak-preserving case, where the term $(\bar \ell  \gamma^\mu P_L\nu _{ \ell })W_\mu^-$ in Eq.~\eqref{eq:int} vanishes. Lastly, since the leptonic decays of charged mesons in the SM are two body, their rates are helicity suppressed. This results in relative enhancements in the branching ratios to ALPs by ratios of meson to lepton masses or phase space factors. 

\begin{figure*} 
\includegraphics[width=\textwidth]{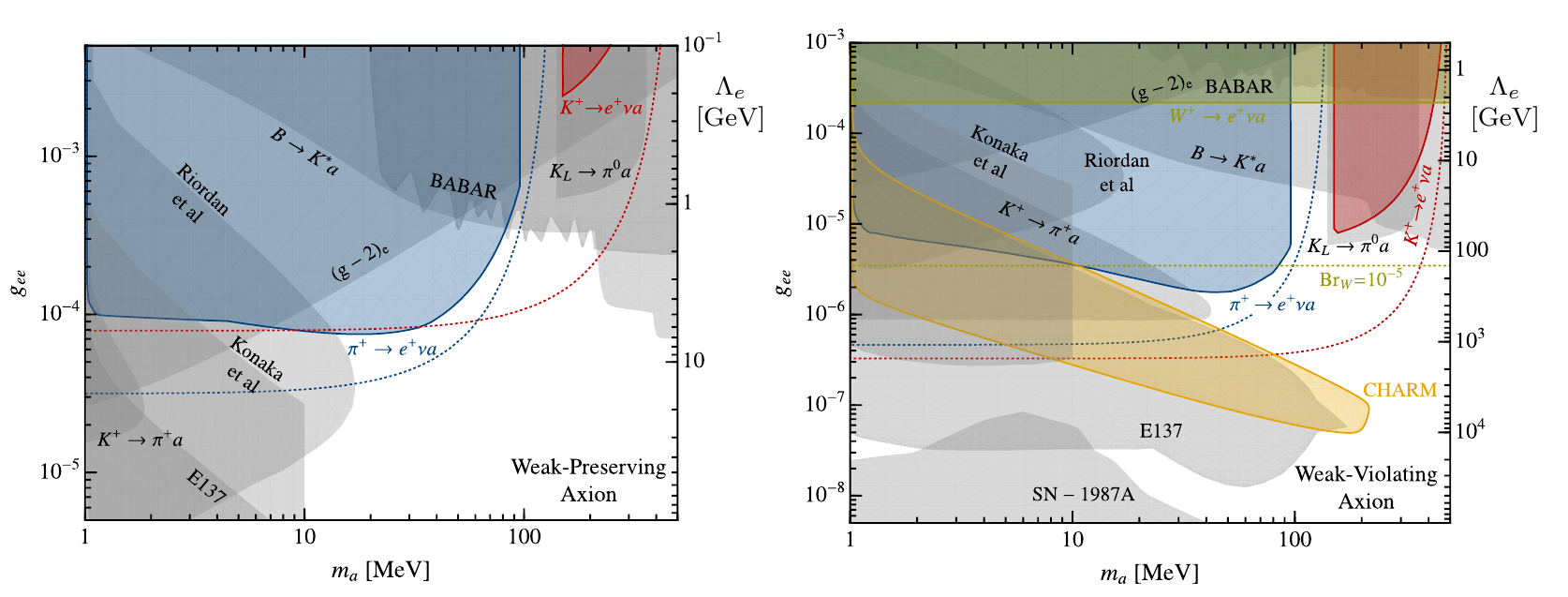}
\caption{The bounds on the coupling $ g _{ e e } $ of leptophilic ALPs interacting with electrons. For comparison, we also plot $ \Lambda _e \equiv m _e / g _{ e e  } $ on the right axes. The gray regions are ruled out by know constraints from the BaBar experiment~\cite{BaBar:2014zli}, searches in electron beam dumps~\cite{Konaka:1986cb,Riordan:1987aw,Bjorken:1988as,Bross:1989mp}, electron $ g-2 $~\cite{Morel:2020dww,Parker:2018vye}, supernova~\cite{Lucente:2021hbp}, and meson FCNCs~\cite{KTeV:2003sls,LHCb:2015ycz,NA62:2020xlg}. Searches for leptonic charged meson decays are shown in blue ({\color{c1} {\bf pions}})~\cite{SINDRUM:1986klz} and red ({\color{c3} {\bf kaons}})~\cite{Poblaguev:2002ug}, while searches for rare {\color{c6} {\bf $ W $ boson}} decays are shown in olive. Dashed lines indicate a rough potential sensitivity with dedicated searches at the PIONEER experiment~\cite{PIONEER:2022yag}, kaon factories~\cite{Goudzovski:2022vbt}, and at the Large Hadron Collider, as described in the main text. Constraints from leptonic charged meson decays in the {\color{c2} {\bf CHARM}} proton beam dump are shown in yellow \cite{CHARM:1985anb}. {\bf Left:} A weak-preserving interaction where only the right-handed electron interacts with the ALP (model $ {\rm WP} $). {\bf Right:} A weak-violating interaction of a pseudoscalar ALP (model WV).}
 \label{fig:geebounds}
\end{figure*}

Various experiments have sensitivity to ALPs from leptonic charged meson decays. For brevity, we consider only the case of ALPs interacting with electrons and electron neutrinos ($ \ell = e $), though our study can be easily generalized to include couplings to muons and taus. We also assume the absence of additional decay modes of the ALP, e.g. into photons, which would reduce the considered signal rates that assume the ALP decays $100\%$ to electrons. To be concrete, we make projections for two models with leptophilic interactions:
$$
\begin{array}{lrl}
\text{Weak-Violating}: & \displaystyle j ^\mu_\text{PQ} = & \displaystyle \frac{  g _{ e e } }{ 2 m _e } \bar{e} \gamma ^\mu \gamma _5 e ~, \\ 
\text{Weak-Preserving}: & \displaystyle j ^\mu_\text{PQ} = & \displaystyle \frac{ g _{ee} }{ m _e } \bar{e} \gamma ^\mu P _R e ~,
\end{array} 
$$
which we label WV and $ {\rm WP} $. The two cases correspond to $\bar g_{\ell\ell}=g_{\nu_\ell}=0,~g_{\ell\ell}=g_{ee}$ and $g_{\nu_\ell}=0,~ \bar g_{\ell\ell}=g_{\ell\ell}=g_{ee}$ in Eq.~\eqref{eq:Jint}, respectively. A weak-preserving model with couplings to left-handed electrons and neutrinos gives results very similar to the weak-preserving case considered here. The normalization is chosen such that the Yukawa-like term is of the form commonly found in the literature, $ g _{ e e   } a \bar{e} \gamma _5 e $.  In the limit $m_a\ll m_\pi$, we find 
 \begin{equation} 
 \frac{\mathcal{B} ( \pi ^+   \to e ^+   \nu a ) }{10^{-11}} \simeq \left\{  \begin{array}{lc} \left( \displaystyle \frac{g_{ee}}{4.6 \cdot 10^{-7}} \right)^2 & {\rm WV}\\ \displaystyle \left( \frac{g_{ee}}{3.1 \cdot 10^{-5}} \right)^2 & {\rm WP}  \end{array} \right. \hspace{-0.1cm}
 . 
\end{equation} 
As expected, the branching ratio in the weak-violating case is enhanced compared to the weak-preserving case. The SINDRUM experiment searched for $e^+e^-$ resonances in $ \pi ^+ \rightarrow e ^+ \nu a, a \rightarrow e ^+ e ^- $ with sensitivity to branching ratios $ {\cal O} ( 10 ^{ - 10} ) $~\cite{SINDRUM:1986klz}. We used the limits given as a function of ALP mass and lifetime to generate the bounds shown in blue in Fig.~\ref{fig:geebounds}. More recently, the PIENU collaboration searched for $ \pi ^+  \rightarrow e ^+  \nu a  $ with the axion remaining invisible~\cite{PIENU:2021clt}, but we find the measurement is not yet able to probe unexplored parameter space. Charged pion decays into visible final states can be searched for at the proposed pion-at-rest experiment PIONEER~\cite{PIONEER:2022yag}. While the capabilities of the experiment are still being determined, the final phases of the experiment plan on reaching a pion flux of over 10 MHz and might have sensitivity to branching ratios of $ {\cal O} ( 10 ^{  - 11} ) $ with under one day of continuous running. As a benchmark, we show a branching ratio of $ 10 ^{ - 11} $ in Fig.~\ref{fig:geebounds} (dashed blue). 

Leptonic charged kaon decays can also produce ALPs with approximate rates (again in the limit $m_a\ll m_K$),
 \begin{equation} 
 \frac{\mathcal{B}(K ^+  \to e ^+  \nu a)}{10^{-10}} \simeq \left\{  \begin{array}{lc} \left( \displaystyle \frac{g_{ee}}{3.3 \cdot 10^{-7}} \right)^2 & {\rm WV}\\ \left(\displaystyle  \frac{g_{ee}}{7.9 \cdot 10^{-5}} \right)^2 & {\rm WP}  \end{array} \right.
 . 
\end{equation} 
While there is no dedicated analysis for this decay mode, E865 searched for the SM process $ K ^+ \rightarrow e ^+ \nu e ^+ e ^- $ finding a branching ratio, $ \mathcal{B} ( K ^+  \rightarrow e ^+ \nu e ^+ e ^- ) = ( 248 \pm 20 ) \times 10 ^{ - 10} $ when the invariant mass of two electrons exceeded 150~MeV~\cite{Poblaguev:2002ug}. We bound the ALP parameter space by requiring the ALP channel does not exceed twice the uncertainty in this measurement. Dedicated searches at kaon factories, which are currently reaching branching ratios below $ {\cal O} ( 10 ^{ - 10} ) $ (see, e.g., the search for $ K ^+  \rightarrow \pi ^+ \nu \nu $~\cite{NA62:2020fhy}), could significantly improve this bound~\cite{Goudzovski:2022vbt}. We show the existing bound as the red region in Fig.~\ref{fig:geebounds} as well as a reference branching ratio of $ 10 ^{ - 10} $ as the dashed line. 

Leptophilic ALPs can similarly be searched for in heavier meson decays such as $ D ^+ \rightarrow e ^+ \nu a $, $ D _s ^+ \rightarrow e ^+ \nu a $, and $ B ^+ \rightarrow e ^+ \nu a $. While weak-violating ALPs have production rates that are more strongly enhanced for heavier mesons, we estimate that with the relatively inefficient production of such mesons in experiments, current measurements of $D$ and $B$ mesons cannot compete with other bounds.

ALPs produced in leptonic decays of all mesons can be relevant for proton beam dump experiments, which produce sizable numbers of charged mesons. The existing literature on using hadron decays at proton beam dumps to search for hidden sectors focuses on flavor-changing neutral current (FCNC) meson decays such as $ B \rightarrow K a $ and $ K \rightarrow \pi a $~\cite{Bezrukov:2009yw,Dolan:2014ska,Izaguirre:2016dfi,Winkler:2018qyg,Dobrich:2018jyi,Egana-Ugrinovic:2019wzj}. To our knowledge, there have yet to be studies on the prospects of using leptonic decays of charged mesons to probe ALPs at proton beam dumps. For concreteness, we focus on CHARM~\cite{CHARM:1985anb}, an experiment that ran a $ 400 ~{\rm GeV} $ beam into a copper target, followed by a 480~m long dump and by a detector placed 35~m after the dump~\footnote{We follow the previous literature~\cite{Dobrich:2018jyi,Egana-Ugrinovic:2019wzj} assuming a (geometric acceptance, mean energy) of the mesons of (0.03, $ 25 ~{\rm GeV}$) for $ \pi ^+ $ and $ K ^+ $, (0.005, $50 ~{\rm GeV}$) for $ D ^+ $ and $ D ^+ _s $, and (0.005, $ 75 ~{\rm GeV} $) for $  B ^+  $ mesons. Furthermore, we take a hadron absorption length of $ 15.3~{\rm cm} $ (which influences the number of ALPs that reach the detector from pion and kaon decays). We take the reconstruction efficiency of electrons and positrons to be $ 0.5 $\cite{CHARM:1985anb}.}. We estimate the meson production cross-section using Refs.~\cite{Aguilar-Benitez:1991hzq,LEBC-EHS:1987evz}. The final bounds from CHARM are only significant for weak-violating ALPs and are depicted in yellow in Fig.~\ref{fig:geebounds} ({\bf Right}).

\section{\boldmath \texorpdfstring{$ W ^+  $}{W+} boson decay}
\label{sec:Wboson}
ALPs can be radiated in leptonic $ W $ boson decays resulting in $ W ^+ \rightarrow \ell  ^+ \nu _\ell a $. The decay rate is negligible for weak-preserving ALPs but is relatively enhanced by $ m _W ^2 / m _\ell  ^2 $ for weak-violating ALPs, making it a powerful probe of leptophilic ALPs. The channel is particularly important at ALP masses above $ 10.2 ~{\rm GeV} $, the threshold for production at BaBar~\cite{BaBar:2014zli}, where the only competing bounds are from LEP. This effect has been noted in the context of muonphilic gauge bosons~\cite{Karshenboim:2014tka}.

In the limit where all final state particles can be approximated as massless, and focusing on the weak-violating interaction in Eq.~\ref{eq:int}, the branching ratio is
\begin{equation} 
\frac{ \mathcal{B}( W ^+ \rightarrow \ell  ^+ \nu _\ell a ) }{ \mathcal{B}( W ^+ \rightarrow e ^+ \nu ) } = \frac{ 3 }{ 1024 \pi ^2 }\frac{ m _W ^2  }{ m _\ell ^2  }\left( g _{\ell \ell  } - \bar{g} _{ \ell \ell  } + g _{\nu _\ell } \right)  ^2 \,,
\end{equation} 
where, as for the charged meson decays, we normalized to a well-known two-body process.
To our knowledge, there have not been dedicated searches for this decay mode. As a conservative bound, we require this mode not to contribute at a detectable level to the total $ W $ width $ \Gamma _W = 2.085 \pm 0.042 $~\cite{Workman:2022ynf}. This gives the 95\% confidence interval bound in olive in Fig.~\ref{fig:geebounds} ({\bf Right}). A dedicated analysis may dramatically improve the sensitivity, given that searches for other rare $W$ decays by CMS reach branching ratios of the order of $ {\cal O} ( 10 ^{ - 6} ) $~\cite{CMS:2019vaj,CMS:2020oqe}. As a benchmark, we show a branching ratio of $ 10 ^{ - 5 } $ in Fig.~\ref{fig:geebounds} ({\bf Right}).

\section{Discussion and Applications}
\label{sec:discussion}
We revisited the theory behind leptophilic axion-like particles. Our primary insight is the rewriting of the ALP interactions through Eq.~\eqref{eq:int}, which highlights the role of the enhanced charged current interaction. This separates two distinct classes of ALPs depending on how SM fields are charged under the underlying PQ symmetry: weak-preserving or weak-violating. Weak-violating ALPs arise from both renormalization-group flow and from integrating out new electroweak-charged degrees of freedom. Interestingly, the often considered pseudoscalar interaction for the ALP, $ \partial _\mu a (\bar{\ell }\gamma ^\mu  \gamma _5 \ell ) $, is of weak-violating form. Such ALPs can be produced with amplitudes enhanced by energy/lepton mass, providing a wealth of new experimental opportunities. 

In this context, we derived new bounds on ALP couplings to electrons from leptonic decays of charged pions and kaons, the $W$ boson, and from proton beam dumps. They are summarized in Fig.~\ref{fig:geebounds} and compared to known constraints on leptophilic ALPs from the literature (gray regions), including searches at BaBar~\cite{BaBar:2014zli}, the electron $ g -  2 $~\cite{Morel:2020dww,Parker:2018vye}, electron or muon beam dump experiments~\cite{Konaka:1986cb,Riordan:1987aw,Bjorken:1988as,Bross:1989mp}, supernova~\cite{Lucente:2021hbp}, and FCNC decays of mesons, $K\to\pi a$ and $B \to K^* a$~\cite{KTeV:2003sls,LHCb:2015ycz,NA62:2020xlg,NA62:2021zjw}.\footnote{To calculate the $ B \rightarrow K ^\ast e + e ^- $ bounds, we use the form factors taken from Refs.~\cite{Bharucha:2015bzk,Gubernari:2018wyi,Gubernari:2022hxn}. Furthermore, we follow the prescription outlined in Ref.~\cite{Altmannshofer:2017bsz} to compute the SM prediction and assume a vertex resolution of 30~$\mu$m~\cite{Aaij:2014zzy}.} Other than the FCNC bounds, all these processes rely only on the $ a \bar{\ell} \gamma _5 \ell $ interaction. Details on how we implemented these known constraints are given in the supplemental material. We find that our new bounds constrain large parts of previously unexplored parameter space. The interaction scale probed in the weak-violating case is above the weak scale, while in the weak-preserving case, the bound on the interaction scale is milder but applies more generically.

While we presented a few novel possibilities here, there are additional directions that demand further study. One possibility is to look for ALPs in muon decays, $\mu \to e \nu_\mu \nu_e a$. Our rough estimates suggest existing bounds are weaker than those from charged pion decays, though the situation may change in the future with the high statistics muon sample at the MEG II experiment~\cite{MEGII:2018kmf}. Similarly, the large number of $B$ and $D$ mesons that will be produced at LHCb and Belle II might enable searches for weak-violating ALPs with masses of $\mathcal O(1\,\text{GeV})$. Furthermore, enhanced ALP production for weak-violating ALPs could take place inside astrophysical environments which produce charged mesons, such as supernovae. We leave the study of their impact for future work.

While the focus of our work was on the significance of weak-violation in ALP theories, our results also have significant implications for weak-preserving ALPs. In this case, models that are subject to additional relevant constraints include the QCD axion with a mass around a MeV~\cite{Alves:2017avw,Alves:2020xhf}, a model with an ALP mediating dark matter scattering off electrons as a method to understand the Xenon-1T excess~\cite{Buttazzo:2020vfs}, and a model in which an ALP is responsible for the Atomki excess~\cite{Liu:2021wap}.

\section*{Acknowledgements}
We thank the PIONEER collaboration for the opportunity to present our work at their group meeting and provide useful feedback. We thank A. Thamm and S. Renner for useful correspondence about Ref.~\cite{Bauer:2021mvw}. We thank Felix Kling and Luc Darme for comments on the manuscript. The research of WA is supported by the U.S. Department of Energy grant number DE-SC0010107. The research of JD and SG is supported in part by NSF CAREER grant PHY-1915852. This work was completed at the Aspen Center for Physics, which is supported by National Science Foundation grant PHY-1607611.

\bibliographystyle{JHEP}
\bibliography{refs}

\newpage

\widetext
\begin{center}
   \textbf{\large SUPPLEMENTARY MATERIAL \\[.2cm] ``New Opportunities for Detecting Axion-Lepton Interactions''}\\[.2cm]
  \vspace{0.05in}
  {Wolfgang Altmannshofer, Jeff A. Dror, and Stefania Gori}
\end{center}
\setcounter{equation}{0}
\setcounter{figure}{0}
\setcounter{table}{0}
\setcounter{page}{1}
\setcounter{section}{0}
\makeatletter
\renewcommand{\thesection}{S-\Roman{section}}
\renewcommand{\theequation}{S-\arabic{equation}}
\renewcommand{\thefigure}{S-\arabic{figure}}
\renewcommand{\bibnumfmt}[1]{[S#1]}
\renewcommand{\citenumfont}[1]{#1}
\section{Double differential branching ratio for \texorpdfstring{\boldmath $P ^+  \to \ell ^+  \nu _\ell a$}{P+ -> l+ nu a}}
Our primary bounds arise from leptonic decays of charged mesons, $P^+$, into ALPs. In the main text, we reported the abbreviated expression for the rate in the limit of vanishingly small lepton mass (see Eq.~\eqref{eq:BRpienua}), but to set bounds for finite lepton masses, one needs to integrate over the differential branching ratio numerically. The double differential branching ratio of the decay $P ^+ \to \ell ^+  \nu _\ell a$ as a function of the lepton energy $E_\ell$ and the axion energy $E_a$ is given by
\begin{multline} \label{eq:dBR}
 \frac{d\mathcal{B}(P ^+  \to \ell ^+  \nu _\ell a)}{dE_\ell dE_a} = \frac{\mathcal{B}(P ^+  \to \ell^{\prime +} \nu _{ \ell '})}{(1-m_{\ell^\prime}^2/m_P^2)^2} \frac{1}{\pi^2} \Bigg[ \frac{(g_{\ell\ell}-\bar g_{\ell\ell} + g_{\nu_\ell})^2}{16 m_{\ell^\prime}^2 m_\ell^2} \big(2(m_P - 2 E_\ell)E_a - (m_P-2E_\ell)^2 -m_a^2+m_\ell^2 \big) \\
 + \frac{g_{\ell\ell}^2 m_a^2}{2 m_{\ell^\prime}^2} \frac{m_P^2-3m_P(E_a+E_\ell)+2(E_a+E_\ell)^2}{(m_P^2-2m_P(E_a+E_\ell)+m_\ell^2)^2}\\ + \frac{g_{\ell\ell}(\bar g_{\ell\ell} - g_{\nu_\ell})}{4 m_{\ell^\prime}^2} \frac{m_P^2-m_a^2+m_\ell^2-2m_P(2E_a+E_\ell)+4E_a(E_a+E_\ell)}{m_P^2-2m_P(E_a+E_\ell)+m_\ell^2} \Bigg]~.
\end{multline}
Integrating the energies between the phase space boundaries
\begin{eqnarray}
 E_- < &~E_a~& < E_+ ~,\qquad 
 E_\pm = \frac{1}{2}\left[ m_P - E_\ell \pm \sqrt{E_\ell^2 - m_\ell^2} + \frac{m_a^2 (m_P - E_\ell \mp \sqrt{E_\ell^2 - m_\ell^2})}{m_P^2+m_\ell^2-2m_P E_e} \right]  ~, \\
 m_\ell < &~E_\ell~& < \frac{m_P}{2} \left( 1 - \frac{m_a^2}{m_P^2} + \frac{m_\ell^2}{m_P^2} \right)  ~,
\end{eqnarray}
and expanding in the lepton mass $m_\ell$ one recovers the result reported in Eq.~\eqref{eq:BRpienua}. Eq.~\eqref{eq:dBR} is derived from the Feynman diagrams shown in Fig.~\ref{fig:NPdecay} based on the interaction Lagrangian $\mathcal L_\text{int} = - a \partial_\mu j^\mu_\text{PQ}$ with the divergence of the PQ current given in Eq.~\eqref{eq:int}. The diagram on the right hand side of Fig.~\ref{fig:NPdecay} stems from the four point interaction in the last line of Eq.~\eqref{eq:int} and gives the first line in Eq.~\eqref{eq:dBR}. We explicitly cross checked the calculation using the Lagrangian $\mathcal L_\text{int} = \partial_\mu a j^\mu_\text{PQ}$ with the PQ current given in Eq.~\eqref{eq:Jint} and obtained the same result. Note that in this case the right diagram of Fig.~\ref{fig:NPdecay} is absent, but a diagram in which the ALP is radiated from the neutrino line needs to be included.

\section{Model Building Weak-Violation}
Weak violation often arises in models which have mixing between states that are charged under the electroweak symmetry. In this section, we present two examples of UV complete models with effective weak-violating interactions. We work in two-component notation and at the end, rework the expression into the four-component notation used in Eq.~\eqref{eq:Jint}. 

We first consider a sterile neutrino, $ N $, that has a Dirac partner, $ N ^c $, and carries charge one under a spontaneously broken PQ symmetry at the scale $ f _a $. Below $ f _a $, the relevant pieces of the mass Lagrangian are,
\begin{equation} 
{\cal L} \supset - y H L N ^c -  M e ^{ i a / f _a } N N ^c +{\rm h.c.} \,,
\end{equation} 
where we use $ H $ to denote the SM Higgs field and $ y $ is a new Yukawa coupling. Since the sterile neutrino mass arises through spontaneous symmetry breaking, perturbativity of the mass term puts the constraint, $ M \lesssim 4\pi f _a $. Performing the rotation, $ N \rightarrow e ^{ -i a / f _a } N $ moves the axion interaction into the sterile neutrino kinetic term. After electroweak symmetry breaking, the Yukawa interaction generates a mass between $ \nu _\ell $ and $ N $. Defining $ \theta = y v / \sqrt{ 2 } M $, the mass term is diagonalized with the rotation, $ \nu _\ell \rightarrow \nu_\ell  + \theta N $ and $ N \rightarrow N - \theta \nu _\ell  $ at leading order in $ \theta $. There is one massless state (corresponding to the physical neutrino) and one heavy state with mass $ M $. 

After applying the transformation, the interaction Lagrangian includes the terms,
\begin{equation} 
{\cal L} \supset  \frac{ \theta ^2\partial _\mu a }{ f _a } \nu ^\dagger _\ell \bar{\sigma}^\mu  \nu _\ell  + \theta \left(   - \frac{ M }{ v } h  \nu_\ell  N ^c + \frac{ e  }{ \sqrt{2} s _W  } W _\mu ^+ N ^\dagger \bar{\sigma} ^\mu e _L + \frac{ e }{ 2 c _W s _W } Z _\mu \nu ^\dagger \bar{\sigma} N + {\rm h.c.} \right) \,.
\end{equation}  
Matching on to the Lagrangian in Eq.\eqref{eq:Jint}, we find for the electron,
\begin{equation} 
 g _{ \nu _e  } = \frac{ 2 \theta ^2 m _e } {f _a } = 1.0 \times 10 ^{ - 5} \left( \frac{ \theta }{ 0.1} \right) ^2 \left( \frac{ {\rm GeV} }{ f _a } \right)\,. \label{eq:gnu}
\end{equation} 
Since perturbativity bounds limit $ M \lesssim 4\pi f _a $, Eq.\eqref{eq:gnu} gives an upper bound on $ g _{ \nu _e } $ as a function of $ M $ and $ \theta $.

There has been an extensive literature on searching for light sterile neutrinos in many different final states (see, e.g., Refs.~\cite{PhysRevD.97.075016,Bolton:2019pcu,Batell:2022dpx} for a recent summary). The most conservative constraints arise from electroweak precision tests which do not assume any specific decay channel for the sterile neutrinos. For $ M \gtrsim 400~{\rm MeV} $, these bounds constrain $ \theta  \lesssim 0.08 $~\cite{PhysRevD.97.075016}. For these values of mass and mixing angle, the upper bound on $ g _{\nu _e } $ from Eq.\eqref{eq:gnu} is much less stringent then the weak-violating constraints presented in this paper.

It is similarly possible to build a model that generates substantial $ g _{e e  } $ and $ \bar{ g} _{ e e} $ couplings by introducing a vector-like lepton, $E$, charged under the PQ symmetry which mixes with the SM electron and has a Dirac partner, $E ^c $. The relevant mass Lagrangian is,
\begin{equation} 
{\cal L} \supset - H L \left( y _e e _R ^c + y E ^c \right) -  M e ^{ i a / f _a } E E ^c +{\rm h.c.} \,,
\end{equation}  
where we have introduced the SM contribution to the electron mass explicitly, parameterized by the Yukawa coupling, $ y _e $, and a new Yukawa coupling with the vector-like lepton ($ y $). Similar to the sterile neutrino case, the transformation, $ E \rightarrow e ^{ - i a / f _a }E $, moves the axion interaction onto the lepton kinetic term. Defining $ \delta \equiv y v / \sqrt{2} $ and working in the limit where $ m _e \ll \delta \ll M $, the diagonalization works analogously to neutrino case. Defining a mixing angle, $ \theta \equiv \delta / M $ we diagonalize the mass matrix to first order in $ \theta $ with the transformation, $ e _L \rightarrow e _L + \theta  E $, $ E \rightarrow E - \theta  e _L $. The electron mass remains approximately equal to $ m _e $ and there is one heavy state of mass $ M $.

To leading order in $ \theta $ the relevant terms in the interaction Lagrangian are,
\begin{equation} 
{\cal L} \supset \frac{\theta ^2  \partial _\mu a }{ f _a } e ^{  \dagger } _L \bar{\sigma} ^\mu e _L  + \theta \left( - \frac{ M }{ v}  h e _L E ^c  +  \frac{ e }{ \sqrt{2} s _W }W _\mu ^+ \nu ^\dagger \bar{\sigma} E + \frac{ e }{ 2 c _W s _W } Z _\mu e ^\dagger _L \bar{\sigma} E +{\rm h.c.} \right)  \,.
\end{equation} 
This interaction produces the combination of the vector and axial electron-axion interactions, 
\begin{equation} 
 \bar{ g} _{ ee  } - g _{ e e} = \frac{  \theta ^2 m _e } {f _a } = 5. \times 10 ^{ - 6} \left( \frac{ \theta }{ 1} \right) ^2 \left( \frac{ 100~{\rm GeV} }{ f _a } \right) \,.
\end{equation} 
Since $ M \lesssim 4\pi f _a $, this in turn puts a lower bound on $ f _a $. Unlike for the sterile neutrino, such a heavy fermion is constrained to have a mass $ M $ well above the weak scale to satisfy collider constraints. At such masses the mixing angle, $ \theta $, is effectively unconstrained. While the strength of the weak-violating coupling is constrained to be smaller than in the previous example, beam dump, meson decay, and $ W $ boson decay searches are sensitive enough that they can supersede these direct constraints on the vector-like leptons.

\section{Existing bounds on ALP couplings to leptons}
\label{sec:bounds}
In the main text, we presented new search strategies to discover leptophilic ALPs. Here, we summarize various other bounds on the electron coupling in the literature and cast them in terms of the interactions in Eq.~\eqref{eq:int}. The known bounds on leptophilic ALPs are split into several types: $ e ^+ e ^- $ colliders (Babar), electron $ g -  2 $, electron beam dump experiments, supernova, and searches for FCNC decays of mesons. The electron beam dump~\cite{Konaka:1986cb,Riordan:1987aw,Bjorken:1988as,Bross:1989mp} and supernova~\cite{Lucente:2021hbp} ALP searches used the $ a \bar{\ell} \gamma _5 \ell $ interaction and can be trivially recast on our parameter space. 
Other than the FCNC bounds, the rest of the processes also rely entirely on the $ a \bar{\ell} \gamma _5 \ell $ interaction but need to be recast with more care.

The BaBar experiment searched for dark photons produced through electron-positron annihilation at a center of mass energy of $ \sqrt{s} = 10.58 ~{\rm GeV} $~\cite{BaBar:2014zli}. We recast this search for $ e ^+ e ^- \rightarrow \gamma a$ by matching the total ALP production cross-sections to that of dark photons below the muon threshold. Above this mass, the additional decay channels of the dark photon modify the bound, and we use a plot of the bound on the total cross-section provided in the supplemental material of Ref.~\cite{BaBar:2014zli} to extend the result to higher masses. 

The SM prediction of the electron $ g -2 $ depends on measurements of the fine-structure constant. These measurements have recently had some controversy, with the most precise results using Cesium~\cite{Parker:2018vye} and Rubidium~\cite{Morel:2020dww} disagreeing significantly. Furthermore, the robustness of the bound on ALPs from electron $ g - 2 $ has been questioned in the literature since it was observed that the two-loop contribution could, in principle, cancel the one-loop contribution depending on the relative sign of $ g _{ e e} $ and the axion-photon coupling~\cite{Alves:2017avw} (see also Refs.~\cite{Marciano:2016yhf,Buen-Abad:2021fwq}). To set a limit, we use the less constraining of the two $ g - 2 $ measurements (Cesium) and the 1-loop contribution, but one should note that this constraint has a degree of uncertainty. 

In contrast to the rest of the existing measurements and searches, the FCNC bounds are the only ones that rely on the anomaly terms in Eq.~\eqref{eq:int}. Through additional loops, these terms contribute to flavor changing quark couplings of the ALP due to the weak interactions,
\begin{equation}
 \mathcal L \supset \frac{\partial_\mu a}{m_\ell} g_{d_i d_j} (\bar d_i \gamma^\mu P_L d_j) ~+~ \text{h.c.} ~.
\end{equation}
From Refs.~\cite{Bauer:2020jbp,Bauer:2021mvw} we find at the leading log,
\begin{equation} \label{eq:gsd}
 g_{d_i d_j} \simeq -\frac{g^2}{16 \pi^2} V_{ti}^* V_{tj} \Big[ \frac{g^2}{16 \pi^2} \frac{3}{8} (g_{ee} - \bar g_{ee} - g_{\nu_e}) F(x_t) + \frac{g^{\prime\,4}}{(16 \pi^2)^2} \frac{17}{96} (g_{ee} + \bar g_{ee}) x_t \log^2\left(\frac{\Lambda^2}{m_t^2}\right) \Big]~,
\end{equation}
with $x_t = m_t^2/m_W^2$ and the loop function
\begin{equation}
 F(x) = x \left( \frac{1}{1-x} + \frac{x \log x}{(1-x)^2} \right) ~.
\end{equation}
The first term in~\eqref{eq:gsd} corresponds to a finite 2-loop contribution that uses the $ a W \tilde{W} $ interaction. The second term is the leading 3-loop contribution that uses a combination of axion-neutral current boson interactions. This is the dominant contribution for purely right-handed electrons.  $ \Lambda $ parameterizes the scale at which new states come in to regulate the diagram, and we take it to be 1~TeV.

Using Eq.~\eqref{eq:gsd} we can obtain the various meson decay rates. The phenomenologically relevant ones are~\cite{Bauer:2021mvw},~\footnote{Note that the FCNC decay modes are small in comparison to the charged current leptonic rates and hence are not the dominant driver of the bounds from CHARM discussed in the main text.} 

\begin{eqnarray}
 \Gamma(K^+ \to \pi^+ a)  &=& \frac{|g_{sd}|^2}{64\pi} \frac{m_{K^+}^3}{m_\ell^2} \left(1-\frac{m_{\pi^+}^2}{m_{K^+}^2}\right)^2
 \lambda^{1/2}\left(\frac{m_{\pi^+}^2}{m_{K^+}^2}, \frac{m_a^2}{m_{K^+}^2}\right) ~, \label{eq:Kpl} \\
 \Gamma(K_L \to \pi^0 a)  &=& \frac{\big( \text{Im}(g_{sd})\big)^2}{64\pi} \frac{m_{K^0}^3}{m_\ell^2} \left(1-\frac{m_{\pi^0}^2}{m_{K^0}^2}\right)^2
 \lambda^{1/2}\left(\frac{m_{\pi^0}^2}{m_{K^0}^2}, \frac{m_a^2}{m_{K^0}^2}\right) ~,\label{eq:KL} \\
 \Gamma(B \to K^* a)  &=& \frac{|g_{bs}|^2}{64\pi} \frac{m_{B}^3}{m_\ell^2}
  |A_0^{B\to K^*}(m_a^2)|^2\lambda^{3/2}\left(\frac{m_{K^*}^2}{m_{B}^2}, \frac{m_a^2}{m_{B}^2}\right) ~,\label{eq:BKs}
\end{eqnarray}
where $\lambda(a,b) = 1+ a^2 + b^2 - 2(a + b + ab)$ and we take the form factor $A_0^{B\to K^*}$ from Ref.~\cite{Bharucha:2015bzk} (see also Refs.~\cite{Gubernari:2018wyi,Gubernari:2022hxn}).
We now briefly summarize the different experimental searches. 

The branching ratio of $K^+ \to \pi^+ a$ with a long-lived $a$ decaying to visible SM particles is constrained by searches at NA62~\cite{NA62:2021zjw}. The limits on $\mathcal{B}(K^+ \to \pi^+ a)$ are given in Ref.~\cite{NA62:2021zjw} as a function of the ALP mass and lifetime and can be directly translated into the leptophilic ALP parameter space. 

The process $K_L \to \pi^0 a, a \to e^+e^-$ can be constrained by searches for the rare SM process $K_L \to \pi^0 e^+e^-$. The search at KTeV~\cite{KTeV:2003sls} found $\mathcal{B}(K_L \to \pi^0 e^+e^-) < 2.8\times 10^{-10}$ at 90\% C.L. The analysis applied cuts on the di-electron invariant mass $140 \,\text{MeV} < m_{ee} < 363\,\text{MeV}$ and thus can be used to constrain ALPs in this mass window. Neglecting the SM prediction for $\mathcal{B}(K_L \to \pi^0 e^+e^-)$, as it is an order of magnitude smaller than the experimental bound, and assuming that the analysis selects SM and ALP events with approximately the same efficiency, we apply the experimental bound directly to $\mathcal{B}(K_L \to \pi^0 a) \times \mathcal{B}(a \to e^+e^-)$. 

Finally, the observed differential distribution of $d \mathcal{B}(B\to K^*e^+e^-)/dq^2$ measured by the LHCb collaboration~\cite{LHCb:2015ycz} can be used to set a bound on $ \mathcal{B}(B \rightarrow K^{*} a)$. We follow the procedure outlined in Ref.~\cite{Altmannshofer:2017bsz} to compute the SM predictions in the three lowest $q^2$ bins and compare them with the corresponding measurements. This leads to
\begin{align} 
&\mathcal{B}(B\to K^*e^+e^-)<1.6\times 10^{-7},~q ^2\subset[0.0004,0.05]{\rm{GeV}}^2~, \notag \\
&\mathcal{B}(B\to K^*e^+e^-)<1.6\times 10^{-8},~q ^2\subset[0.05,0.15]{\rm{GeV}}^2~, \notag \\
&\mathcal{B}(B\to K^*e^+e^-)<7.0\times 10^{-9},~q ^2\subset[0.15,0.25]{\rm{GeV}}^2~.
\end{align} 
For the first two bins, we quote the $2\sigma$ bounds, while for the third bin, we set the bound at the $3\sigma$ level. This choice was made to account for the fact that the LHCb measurement in the third bin is low and excludes the SM prediction at $2\sigma$. To set a bound on the ALP parameter space, we compute the branching ratio for $B\to K^*a, ~a\to e^+e^-$, requiring the ALP to have a transverse displacement smaller than the LHCb vertex resolution $\sim 30~\mu$m~\cite{Aaij:2014zzy}, adapting the procedure described in Appendix D.4 of Ref.~\cite{Bauer:2021mvw}. 


\end{document}